\documentclass{evn2004}
\usepackage{txfonts}
\newcommand{\etal}{{\em et al.~}}
\begin{document}
\title{The Australian experience with the PC-EVN  recorder}

\author{R. Dodson\inst{1,2},
S. Tingay\inst{3}, C. West\inst{3}, 
C. Phillips\inst{4}, 
A. Tzioumis\inst{4}, 
J. Ritakari\inst{5}, 
F. Briggs\inst{6}
}
\authorrunning{Dodson \etal}

\institute{Institute of Space and Astronautical Science, 
                 JAXA, 3-1-1 Yoshinodai,
                 Sagamihara, Kanagawa 229-8510, Japan
\and
University of Tasmania, Hobart, 7000
\and
Swinburne University of Technology, PO Box 218, Hawthorn 3122,
  Melbourne, Australia
\and
Australia Telescope National Facility,
                 Commonwealth Scientific and Industrial Research Organization,
                 P. O. Box 76, Epping NSW 2122, Australia
\and
Mets\"ahovi Radio Observatory, Mets\"ahovintie 114, 02540
                 Kylm\"al\"a, Finland 
\and
RSAA, The Australian National University}

\abstract{

We report on our experiences using the Mets\"ahovi Radio Observatory's
(MRO) VLBI Standard Interface (VSI, Whitney 2002) recorder in a number
of astronomical applications. The PC-EVN device is a direct memory
access (DMA) interface which allows 512 megabit per second (Mbps) or
better recording to ``off the shelf'' PC components. We have used this
setup to record at 640 Mbps for a pulsar coherent dispersion system
and at 256 Mpbs for a global VLBI session. We have also demonstrated
recording at 512 Mbps and will shortly form cross correlations between
the CPSR-II and the PC-EVN systems.

}

\maketitle

\section{Introduction}

Astronomy has always battled against the extremely weak signals
received from objects impossibly far away. We can improve the
situation by recording the signals with cooler feeds, observing for
longer, or by collecting wider bandwidths, as expressed by the
radiometer equation (see e.g. Kraus 1986). Cooler feeds are difficult
and telescope time is precious, so the most profitable area for
improvement is to record wider bands. The Long Baseline Array (LBA)
uses a bandwidth of 16MHz, from receivers with a bandpass that is
typically the order of a GigaHertz. This is almost criminally
wasteful.

With modern digital technology it is possible to sample increasingly
wider bandwidths with a higher number of bits. Furthermore by recording
to consumer electronic hardware (the so called ``Commercial off the
Shelf'' or COTS approach (e.g. Whitney 2003)) the equipment is cheap,
reliable and easy to repair, replace and upgrade.

In Australia the VLBI recorders are S2s (plus a few sites with MkIIIs
which will be all upgraded to MkVs soon). The S2 is a
multi-video tape digital recorder (Cannon \etal 1997) which records
128Mb/s. In Australia it is fed by the ATNF Data Acquisition System
(ATNF DAS). We have found that, with very little effort, the DAS can
be configured to drive Mets\"ahovi's PC-EVN card. The maximum data
rate of the DAS is 512 Mb/s which nicely matches the quoted maximum
rate of a single PC-EVN. If we wish to use the flexible digital
filters of the DAS we are, however, constrained to 256 Mb/s.

CSIRO has secured agreement with AARnet to provide access to dark
fibres that pass near their research centres in regional areas. The
network will be equipped with 10~Gbs capable (Dense Wave Division
Multiplexing) infrastructure. The VLBI project is providing the major
scientific impetus for this bandwidth.
As a demonstration system for real time VLBI we have installed a
parallel ``fringe checker'' system where the PC-EVN recorder driven by
the data-out port of the S2. We now can confirm the data integrity of
an experiment during the run.
We are also taking part in the differential VLBI tracking of the Huygens
probe (Gurvits et~al 2004) using this same hardware.

\section{PC-EVN Hardware}

The boards purchased from MRO were their two VSI boards; VSIB (a DMA card)
and VSIC (a converter card). The VSIB card allows data to be written to
or read from the hard drives on the PC.
The provided operating system was debian linux (the stable woody
series with kernel 2.4.19 and the big physical area patch).
The recommended PC motherboard is the K7 series from MSI with an AMD
processor. In Hobart we use these with four IDE disks (200~GBytes in
size) from the motherboard. A boot disk (and CD) is run from a PCI IDE
card, making the total disk capacity one terabyte. The data disks are
mounted in a (software) RAID0 array, see Figure \ref{fig:box}.
Swinburne used the Dell~1600C server machine, because with these
the data disks can be external Apple Xraid disks on a 64~bit PCI
bus. These were run using SuSe~8.2 linux OS. Using alternate
motherboards and OS versions caused no problems.

A third system has just been purchased, which is primarily for the
Huygens experiment, but also is being used as a test bed. This is
based on the K8 motherboard, which has a 1GB ethernet on-board (so not
on the same data bus as the hard drives). This is being used to
investigate direct streaming of the data collected by the VSIB across
the network. It is currently configured with four 400~GB hard drives.

The VSI cards cost 565 Euros each, the PC the order of AU\$2000 and about
the same for four multi~Gbyte hard drives. Therefore the total cost of
the system was around AU\$5500. 

\section{The VLBI system}

   \begin{figure}
   \centering
   \vspace{200pt}
   \includegraphics{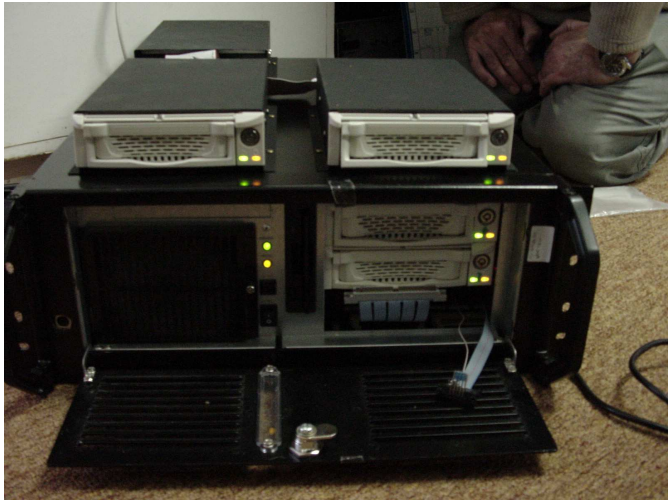}
   \caption{The VLBI system for Ceduna with four demountable 200GB IDE
   drives. The VSIC connector is the 50 way ribbon connector at the
   bottom right.
   \label{fig:box}}
   \end{figure}

The VSIC card converts many ``legacy'' VLBI digital signals into the
ordering and levels (LVDS) of the VSI standard required by the VSIB
DMA card. In the Hobart system the VSIC card is mounted and powered
inside the PC case. It can be seen at the bottom right of the PC shown
in Figure \ref{fig:box}. For the Swinburne systems the VSIC card
was enclosed in its own box. The card will handle up to 32 channels of
data, plus the required signal clocks and 1PPS signals.

The VSIC card will format data with S2 pin outputs (as well as VLBA,
Mk3 and Mk4 formats), so it was a trivial task to connect the VSIC to
the cable normally going to the S2 recorder. The provided program {\bf
wr} was used to collect the data. {\bf wr} reads a minimum of 8
channels, whilst the maximum rate S2 recording modes (32x4-2) records
two IFs of 16 MHz of data at 2 bits into 4 channels. The DAS modes
used (VSOP and MP16S), however, encode four IFs of 16 MHz of data at 2
bits into 8 channels. Therefore we could record double the usual LBA
bandwidth using the PC-EVN system. We have altered {\bf wr} to allow
us to drop subsections of the data read, before writing. These
modes allow longer, and more flexible recording programs. More details
can be found in Dodson (2004).

We have performed (in April 2004) global eVLBI observations. Most
Australian telescopes ({\bf ATCA, Mopra, Parkes, Tidbinbilla} and {\bf
Ceduna}) recorded on PC-EVN machines, and {\bf Hobart} recorded to
their Mk5a system. Also recording on Mk5 systems were {\bf
Hartebeesthoek} (SA) and {\bf Pietown} (USA). Data was recorded to the
K5 system at {\bf Kashima} in Japan. 
The global VLBI data has been transported (by shipping the disks) to
Swinburne, and correlated on their supercomputer. A full report on the
software correlator, the experiments and the results is in
preparation, but phase closure has been shown between the telescopes.

   \begin{figure*}
   \centering
   \vspace{280pt}
   \includegraphics{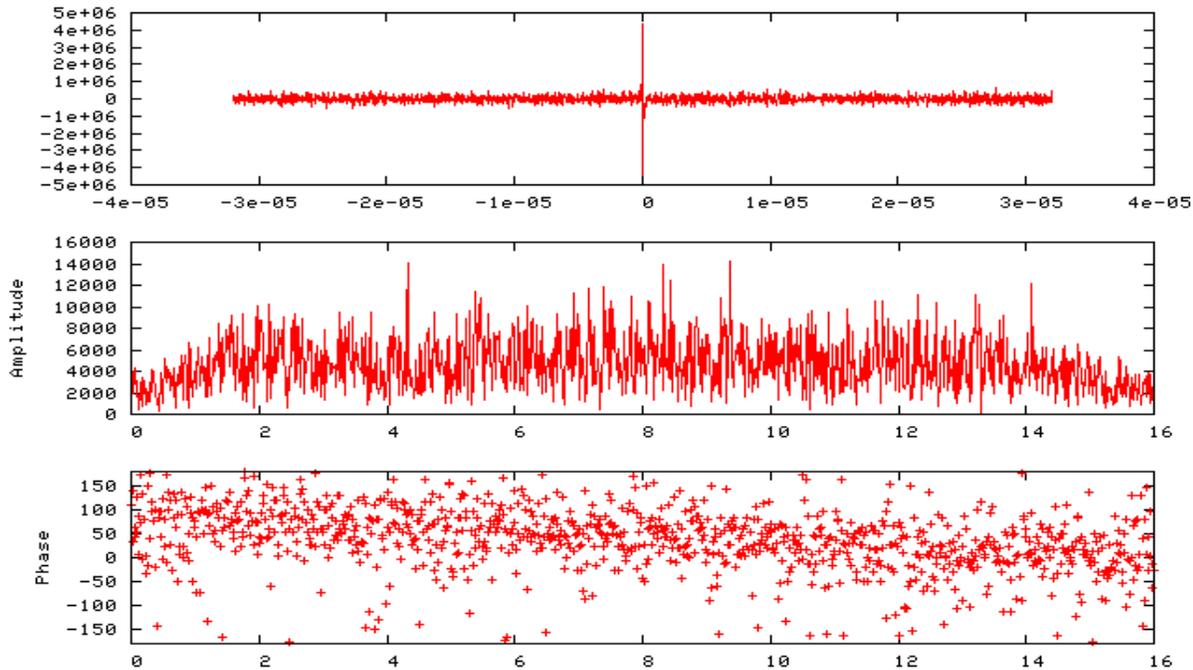}
 \caption{The automatic fringe checkers output from an experiment on
   Sept 12 between ATCA and Parkes. Observing 1104-445 at L band for one second
   with a 16 MHz bandpass.
 \label{fig:fringe}}
    \end{figure*}

During the experiment small sections of the recorded data from Ceduna,
Mopra, Parkes, Tidbinbilla and Narrabri were transfered to various
software correlators to perform near real time fringe checking.
After the success of this (and the `on the fly' discovery of a swapped
polarisation at the ATCA) we set to developing a permanent, parallel
fringe checking system. The S2 system mirrors the data in (on port C1)
on a data out port (C2a). We now record this during regular VLBI
experiments, and immediately provide data integrity confirmation. The
fringe checking experiments are limited to the ATNF antenna at
present, but we have plans to expand this to Tidbinbilla, Hobart and
Ceduna. The data transport requirements are slight; if we
checked one second of data every hour the rate would be less than 5kB/s.

\section{Differential VLBI observations of the Huygens Probe}

JIVE has proposed to use a global array of VLBI antennae to
observe the Huygens probe during its descent into the atmosphere of
Titan on 14 January 2005. One of the key science goals is to measure
the dynamics of Titan's atmosphere.  Phase referenced VLBI will be
used to track the probes position during its descent. Along with the
Doppler shift of the probes signal, the full 3 dimensional velocity of
the probe will be measured.  LBA participation in the experiment will
be important due to the position of the probe on the sky.
Now that Australia has the capability to install 0.5 Gb/s recorders at
all antennae for a few thousand (AU) dollars we are able to commit a
significant number of baselines to this project.

For this experiment two DASs per PC-EVN recorder are required, so
where possible the MkV system is being used freeing up resources for
other antennae.
While an individual DAS can provide two 64~MHz bands the digital
filters are limited to providing four narrower (16~MHz) bands (from
within the input 64~MHz bands). As observations need to be spread over
a large frequency range for the atmospheric calibration the only
solution is to have two separate systems, one for the probe signal,
and one for the spread frequency calibration signal.
Combining the signals from two (independent) DAS, driven by the clock
from one of them, is an issue but has been demonstrated within the
last few days.

\section{Other Projects}

The Mets\"ahovi recorder, being a bare-bones, and therefore very
flexible, system has led us to use it for a number of other
projects. These required a AD converter, buffered out as
LVDS levels into the VSIB. 

\subsection{The Mt. Pleasant Pulsar backend}


Hobart has been monitoring the Vela pulsar's pulse time of arrival
(TOA) for over twenty years. The original system (which caught a
glitch in the first week of operation (McCulloch \etal 1983)) has been
upgraded over the years to one that;

\begin{itemize}

\item monitors Vela for 18 hours a day (and a second glitching pulsar
  for the remaining 6 hours).

\item collects two minute averaged profiles at three frequencies (635,
  990 and 1390~MHz). The bandwidth is matched to the dispersion time
  (for a dispersion measure of 69~pc~cm$^{-3}$). The automatically
  generated fit to these data is available on the
  web\footnote{http://www-ra.phys.utas.edu.au/$\sim$rdodson/0835-4510.par}.

\item collects continuously sampled data at 990~MHz (incoherently
  dedispersed to increase the bandwidth by a factor of eight). From
  this single pulse system 10 second profiles are constantly formed
  and checked for the occurrence of a glitch. These data are saved for
  3 days for deep analysis of immediate post (and pre) glitch
  behaviour (Dodson, McCulloch \& Lewis 2002).

\item now also collects continous sampled data at 635~MHz across a
  25~MHz bandpass offset from baseband by 10~MHz, increasing the
  bandwidth one hundred fold. This data is buffered for 2.5 hours,
  after which it is overwritten. Once a glitch is detected by the
  single pulse system the data is saved for off-line coherent
  dedispersion and profile forming.

\end{itemize}

%
Our new system produces TOA's with accuracy of the order of 0.1
msec every second (as opposed to every 10 seconds with the single
pulse or 120 seconds with the multi-frequency systems).

The IF (baseband to 40 MHz) was fed in to a MAXIM 1448 A/D mounted on
an evaluation card (AU\$330 each), along with a 80 MHz clock. Two of
these provide the two polarisations. The output of these are TTL, so
the four most significant bits (of the 10 bits provided by the MAX1448
card) are buffered to LVDS along with the clock signal, and combined
with an external 1PPS signal. These are fed directly into the
VSIB. 
%
In running the VSIB at 80~MHz we are clocking the data at two and a
half times the VSI standard, nevertheless this produced no
complications.
The recording software is that provided ({\bf wr}), with minor
modifications to allow continous looped recording and shared memory
control. This system is now running and recording data. However it
missed the last glitch (Dodson \etal 2004) as we switched to a
debugging mode 24 minutes before the event (Murphy per. comm.). 

\subsection{The Stromlo Streamer}

For radio astronomers, the radio frequency interference (RFI)
environment continues to get worse, and has become a critically
important issue in the development of the next generation of radio
telescopes, including the Square Kilometre Array. 
%
For the SKA site evaluations and interference mitigation
investigations Mount Stromlo Observatory (Australian National
University) has used a near identical setup to the pulsar system as a
flexible test bed for emulating radio astronomy backends in
software. The system is portable, carrying with it a low-precision 64
MHz clock and 4 MAXIM 1448 A/D cards for accepting 4 IFs as input. The
flexibility provided by the MRO {\bf wr} routine allows the A/Ds to
deliver different numbers of bits precision and different sampling
rates without reconfiguring the hardware.
The most often used application has been in the area of RFI mitigation
to further the program described in Briggs, Bell and Kesteven (2000).


\subsection{A mobile VLBI recorder}

We are collaborating with Auckland University of Technology to build a
VLBI recorder for use with a satellite tracking dish. This is a
demonstration project for future, large scale, VLBI between Australia
and New Zealand. The recorder is a Dell system, writing to an Xraid
disk pack. The time standard is a GPS disciplined clock. Again we are
going to use the MAX1448 AD to sample the analogue baseband signal,
and will record two 16~MHz bands for correlation with antennae in
Australia. This baseline fills a significant gap between
Hartebeesthoek and the Australian LBA baselines.

\section{The future}

Using the PC-EVN systems in place we will collect the data from the S2
formatter output port (C2a), and fringe check the data on the fly
during all VLBI observations. Automation of the data collection and
checking is being fine tuned. This will allow rapid checking and
debugging of the array and will further improve the success rate of
VLBI observations

The ATNF expect to have a 10~Gb/s network installed at all their
telescope sites by early 2005. We will, therefore, be able to
quickly catch up with the current e-VLBI systems of the Europe, Japan
and the USA. The project time line is to have 0.5 Gb/s recording
available at all antennae by the mid 2005. This will be recorded to
removable disks and correlated in the Swinburne software correlator
(West 2004). The new ATCA broadband backend (sited at Narrabri) is
projected to be completed in 2007. It will be able to correlate the
VLBI-station data, delivered over fibre, in real time. The correlator
will be free as during the VLBI runs the ATCA operates as a
VLBI-station itself.

\begin{figure}
   \centering
   \vspace{227pt}
   \includegraphics{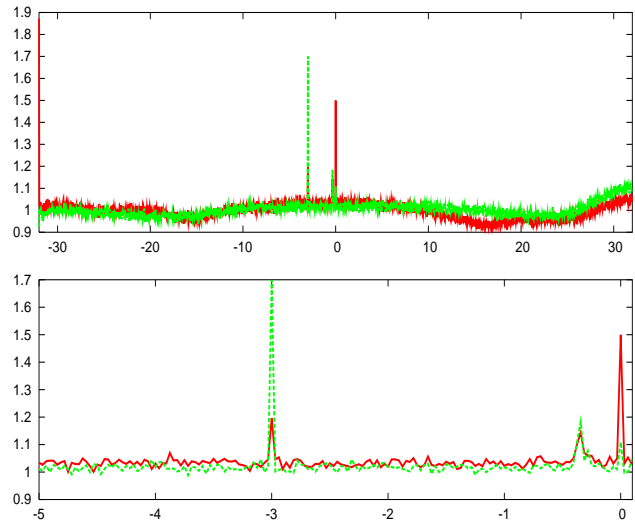}
   \caption{Autocorrelation of 0.7 seconds of 64~MHz bandwidth data
   observing the maser G309.9+0.5. Both polarisations have been
   recorded and a test tone has been included at 3~MHz below band
   centre (which is 6667 MHz). The bandpass fed into the DAS is not
   very sharp and the aliased edges folded back into the 64~MHz
   causing the rise in the system temperatures at $\pm32$~MHz.
   \label{fig:g309-64}}
   \end{figure}

As a demonstration we have collected the data from the DAS correlator
port rather than the S2 connector port, which allows us to record 2
IFs of 64 MHz (at 2 bits). Figure \ref{fig:g309-64} shows the bandpass
collected at Hobart looking a the methanol maser G309.9+0.5. There is
also a test tone 3 MHz below the band centre.

At Parkes we collected coincident data recorded on the CPSR-II system
to produce the zero baseline cross correlations. We will shortly
attempt to correlate CPSR-II data (from Parkes) with data from two
PC-EVN machines (at ATCA) to demonstrate 1~Gbps VLBI in Australia.
%


The recording rate limitations are from that sustainable by the hard
drives (about 800~Mbs for the slowest system), however even when
greater speeds are possible (with Serial IDE drives or just general
improvements) we will be at the PCI maximum rate. Already we'd like to
be able to read and write over the same bus at the same time, in which
case developing the VSIB card to run on the faster, emerging,
technologies would be of great priority. The estimated cost for this
is 3 man months, and if sufficient interest is generated this will be
undertaken at Mets\"ahovi.

\section{Conclusions}

We have demonstrated the adaptability of the PC-EVN system, and its
low cost of setup in both hardware and manpower. We have used it for
analog RF recording systems and as a digital recorder for the
Australian VLBI system. We will use it as the basis for a fringe
checker system, and as the wide bandwidth recorder for the Huygens
experiment.
We expect to continue to find new and varied uses for these cards.



\end{document}